\documentstyle[preprint,aps]{revtex}

\newcommand{\be}{\begin{equation}}
\newcommand{\ee}{\end{equation}}
\newcommand{\br}{\begin{eqnarray}}
\newcommand{\er}{\end{eqnarray}}
\newcommand{\ds}{\displaystyle}

\def\b{{\beta}}
\def\g{{\gamma}}

\def\d{{\delta}}

\def\o{{\omega}}

\begin{document}
\draft
\title{
$q$-DEFORMED FERMION OSCILLATORS, ZERO-POINT\\
ENERGY AND INCLUSION-EXCLUSION PRINCIPLE
}
\author{P. Narayana Swamy}
\address{
Physics  Department, Southern Illinois University,
Edwardsville, IL 62026, USA}
\maketitle

\begin{abstract}
The theory of Fermion oscillators has two essential ingredients:
zero-point energy and Pauli exclusion principle.
We devlop the theory of the statistical mechanics of generalized $q$-deformed
 Fermion oscillator algebra with inclusion principle ($i.e.,$ without the
exclusion principle), which corresponds to ordinary fermions with Pauli
exclusion principle in the classical limit $q \rightarrow 1$. Some of the
remarkable properties of this theory play a crucial role in the understanding
of the $q$-deformed Fermions. We show that if we insist on the weak
exclusion principle, then the theory has the expected low temperature
limit as well as the correct classical $q$-limit.
\end{abstract}

\vspace{0.5cm}

\noindent PACS numbers: 05.30.-d,\ \ 05.70.-Cd, \ \ 05.70.-Fh, \ \ 05.90.+m\\
\noindent Preprint SIUE/HEP-1/ 1999

\vspace{0.2in}

Let us begin with the classical system of ordinary quantum Fermion
 harmonic oscillators with the spectrum 
\be
E_n = (n - \frac{1}{2}) \hbar \omega,  \quad n=0,1\; ,
\ee
and the Partition function given by
\be
{\cal Z}=\sum_0^1 e^{\b \;  E_n}\;  2 \cosh \frac{\b \hbar \omega}{2},
\label{2}
\ee
where $\b = 1/ kT$ and $k$ is the Boltzmann constant. The free energy is
\be
{\cal F} = -\frac{1}{\b}\ln {\cal Z}= -\frac{1}{\b}\left ( \ln 2 +
\ln \cosh \frac{\b \hbar \o }{2} \right ),
\ee
from which we obtain the entropy:
\be
S = \frac{\partial {\cal F}}{\partial T}= -k \b {\cal F} -  \b \frac{\hbar
\o}{2} \tanh \frac{\b \hbar \o}{2}.
\ee
The internal energy of the quantum Fermion oscillators is then determined
 by
\be
U = {\cal F} + T S = - \frac{1}{2} \hbar \o \tanh \frac{\b \hbar \o}{2}.
\ee
We may also express the internal energy in the form
\be
U= \frac{1}{2} \hbar \o \; \frac{1- e^x}{1+  e^x }=  \hbar \o \left (
 -\frac{1}{2} + \frac{1}{e^x +1}  \right )
\ee
where $x= \b \hbar \o$.  Here the first term contains the zero point energy
 and the second is determined by the Fermi distribution. For $N$
  non-interacting Fermion oscillators, we have $U= N \hbar \o f$
   where $f$ is the probability distribution function and therefore
    we infer that
 \be
f= -\frac{1}{2}+  \frac{1}{e^x+1} =  -\frac{1}{2} \tanh \frac{\b \hbar \o}{2}.
\label{7}
\ee
Alternatively we can derive the form of the distribution  function
 as follows. The occupational probability is
 \be
P_n= \frac{e^{\ds -\b E_n}} {\cal Z} = \frac{e^{\ds -\b E_n }}{2 \cosh 
\frac{\ds \b E}{\ds 2}}\; , \qquad E_n=(n- \frac{1}{2}) \hbar \o, \quad n=0,1.
\ee
Thus we have
\be
f= \sum_0^1 (n-\frac{1}{2})  P_n = -\frac{1}{2} \tanh \frac{\b \hbar \o}{2}
\ee
which can also be expressed as in Eq.(\ref{7}). We may examine the high
temperature limit of the internal energy when $\b \hbar \o << 1$  and 
thus
\be
U_{T \rightarrow\infty} = \lim N \hbar \o  \left (
-\frac{1}{2}+ \frac{1}{e^x + 1} \right )
= \lim N\hbar \o \left (
-  \frac{1}{2}+  \frac{1}
{2 + \b \hbar \o + \frac{1}{2} (\b \hbar \o  )^2 \cdots }
\right ) = 0,
\ee
as expected: the energy of the Fermion oscillator indeed vanishes in the
classical limit, it is purely a quantum effect, when Pauli exclusion
 principle prevails. Next, we may consider the low temperature limit,
  when $\b \hbar \o >> 1$. We then find
\be
U_{T \rightarrow 0}=  \lim N \hbar \o \left (- \frac{1}{2}+ e^{\ds
 - \b \hbar \o} \right )  = -
\frac{1}{2} N \hbar \o
\ee
again as expected.  In this limit, we of course expect the internal
 energy to reduce to the zero point energy, a pure quantum effect.

The standard Fermions are described by the algebra of creation and
annihilation operators defined by
\be
aa^{\dagger}+a^{\dagger}a=1, \quad [N,a]=-a, \quad
[N,a^{\dagger}]=a^{\dagger}, \quad a^2 =(a^{\dagger})^2=0,
\label{12}
\ee
in accordance with Pauli exclusion principle, where $ N $ is the number 
operator. We have suppressed the quantum number indices of the creation and
annihilation operators. The Hamiltonian is given by
\be
H=\frac{1}{2} \hbar \o (a^{\dagger}a-aa^{\dagger})=\hbar \o(N-\frac{1}{2}) 
\ee
with the eigenvalues $E=\hbar \o (N - \frac{1}{2})$. The Fock states
 constructed by
\be
a |n>\, =\sqrt{n}|n-1>, \quad a^{\dagger} |n> =\sqrt{n+1} |n+1>,
\ee
obey the Pauli exclusion principle, and Eq.(\ref{12}) implies $N^2=
 a^{\dagger} (1- a^{\dagger}a )a = N, \quad N(N-1) =0, $  so that the
  number operator has $n=0,1$ as the only allowed eigenvalues.

The conventional theory of $q$-deformed Fermion oscillators consists
 of a straightforward extension of the ordinary Fermion oscillators,
  employing the $q$-analog \cite{Macfarlane,Bied}  of the ordinary
   Fermion algebra. This version is defined by
\be
aa^{\dagger}+ q  a^{\dagger}a=q^N, \quad [N,a]=-a, \quad
[N,a^{\dagger}]=a^{\dagger}, \quad a^2 =(a^{\dagger})^2=0.
\ee
It is well-known that this version of $q$-Fermions  such as in the
work of Hayashi \cite{Hayashi}  and others reduces trivially to the
ordinary undeformed Fermions with Pauli exclusion principle as it
can be shown that the deformation can be transformed away. This
has been demonstrated by R. Parthasarathy et al \cite{Partha} and
by Jing and Xu \cite{Jing}. Accordingly, we shall therefore
investigate the theory of the generalized $q$-deformed Fermions
proposed by Parthasarathy et al \cite{Partha}. This particular theory
allows many Fermion states with inclusion principle,  {\it i.e., }
without exclusion principle, but reduces to the ordinary Fermions with
exclusion principle in the classical limit $q \rightarrow 1$. We shall
present a brief review of this generalized theory and point out some
extraordinary features of this theory before investigating the statistical
mechanics of this generalized theory.

The $q$-deformed theory of generalized Fermions is described by the algebra
of the creation and annihilation operators specified by
\be
b b^{\dagger} + q b^{\dagger} b= q^{-N}, \quad [N, b]=-b, \quad
[N,b^{\dagger}]
=b^{\dagger},
\ee
where $N$ is the number operator, $N \not= b^{\dagger} b$.  The Fock states
are constructed by
\be
N|n>= n|n>, \quad b|n>= \sqrt{[n]} |n-1>, \quad 
b^{\dagger}|n>=\sqrt{[n+1]} |n+1>.
\ee
Thus we find $b^{\dagger}b= [N],  \;  bb^{\dagger}= [N+1], \quad
b^{\dagger}b|n>= [n] |n>, \quad   bb^{\dagger}|n>=[n+1] |n> $ where the
basic (analog of) numbers $[n]= [n]_F $  are defined by 
\be
[n]= \frac{q} {1+q^2}\left \{ q^{-n}- (-1)^n q^n\right \}\; = \;
\frac{q}{1+q^2}\left \{ q^{-n}- (-q)^n \right \}.
\ee
This definition is of fundamental importance in this generalized $q$-deformed
Fermion theory and represents the Fermion basic (analog of ordinary)
numbers. It should be stressed that $[n]$ defined in this manner is
quite different from the case of bosons, with $[n]_B$ \cite{pns,pns1}.
It must be pointed out that this set of basic (analog of)
 numbers is also quite different from the historically well-known
 basic (analog of) numbers  from the work of E. Heine (1847, 1989),
 F.H. Jackson and others \cite{Hexton}.
We shall drop the subscript $F$ (Fermions) henceforward unless the
circumstance requires us to distinguish it from the boson case $[n]_B$ in
the analysis. We have scaled the deformation parameter differently
from that of ref. \cite{Partha}.  Our $q$ is the same as $\sqrt{q}$
of ref. (\cite{Partha}), which simplifies the notation and ought not to
make any difference in the conclusions. The Fock states constructed by
\be
|n>= \frac{1}{\sqrt{[n] !}} (b^{\dagger})^n |0>
\label{18}
\ee
where the factorial is defined by
\be
[n]! = [n]\, [n-1]\, \cdots [1]\; ,
\ee
do not in general satisfy Pauli exclusion principle, and allow
the eigenvalues $n=0,1, \cdots \infty $. The maximum value need not
be $n= \infty$,  could be finite in some cases determined by the value of
$q$ as described in ref. \cite{Partha}. The states constructed in this
manner have some remarkable properties.  Firstly, we note that $[n] =n$
only for $n=0,1$ and $[n] \not=n   $ for $n > 1$,  which result is true
for any $q$. This property is quite different from the case of $[n]_B$.
Secondly, in the limit $q \rightarrow 1$, we observe that
\be
[n]_{q=1} = \frac{1}{2} \{1- (-1)^n \}\;, 
\ee
which takes the values $0, 1$ for even and odd $n$ respectively and
hence in the limit, it is $0,1,0,1,  \cdots $ for $n=0,1,2,\cdots \infty$.
This is in contrast to the case of the $q$-deformed bosons where $[n]_B$
reduces to $n$ in the limit $q \rightarrow 1$  for any $n$.  Thirdly,
by introducing $q=e^{\g}$ we may also express $[n]$ as
\be
[n]= 0,1, \,  -\frac{\sinh 2\g}{\cosh \g}, \,   \frac{\cosh 3\g}{\cosh \g},
\, -\frac{\sinh 4\g}{\cosh \g}, \,  \frac{\cosh 5\g}{\cosh \g}, \,  \cdots
\ee
for $n=0,  1,  2,  \cdots,  \infty$ which result is also quite different
from the case of $q$-bosons \cite{pns}. 
 
Next we see that the classical expression for the Hamiltonian,
$H=\frac{1}{2} \hbar \o (b^{\dagger}b -bb^{\dagger}) $ yields the
$q$-analog of the Hamiltonian of the $q$-deformed theory
\be
H= \frac{1}{2} \hbar \o \left \{   [N] - [N+1]  \right \}.
\ee
The eigenvalues are given by $E= \frac{1}{2} \hbar \o ( [n]- [n+1] )$
and the levels are no longer equally spaced.  To examine the $q \rightarrow
1$ limit, we observe that
\be
\lim_{q \rightarrow 1 }  [n+1]= \frac{1}{2}\left  \{ 1- (-1)^{n+1} \right \}
\ee
which shows that in this limit, the spectrum $E= -\frac{1}{2} \hbar \o,
\frac{1}{2} \hbar \o, \frac{1}{2} \hbar \o, \cdots $  maps on to the
undeformed Fermion energy spectrum $-\frac{1}{2} \hbar \o, \frac{1}{2}
\hbar \o$,   albeit the fact that $n$ is not restricted to $0,1$. Thus
in the classical limit the $q$-deformed generalized Fermion theory
has the same spectrum as the ordinary Fermion oscillators with Pauli
exclusion principle. The fact that Pauli principle prevails in the
limit $q \rightarrow 1$, although the generalized $q$-deformed Fermion
 theory has no Pauli principle ($i.e.,$  obeys inclusion principle),
    can be seen more directly from Eq.(\ref{18}).  We note that $[n]!
=  [n] \, [n-1]\, [n-2]\, \cdots [1]$ reduces in the classical limit
 to $[n]!= 0$ for $n > 1$ and $[n]!=1$ for $n=0,1$ and hence
  $(b^{\dagger})^n|0> =0$ for $n >1$. This may be referred to
  \cite{Partha} as the weak exclusion principle or the exclusion
 principle in the operator sense. This is the fourth remarkable property
 and we must return to it later.

With these preliminaries established,  let us now proceed with the
investigation of  the statistical mechanics of the generalized
 $q$-deformed Fermions. The occupational probability is
\be
P_n= \frac{e^{\ds - \b E_n}}{\cal Z}= \frac{1}{\cal Z} e^{ - \ds \frac{x}{2}}   
(\, [n]-[n+1]\, ),
\ee
where the Partition function is now given by
\be
{\cal Z}= \sum_0^{\infty}e^{-\b E_n}.
\ee
We would be interested in the lowest order iteration of the theory in which
case we may employ the lowest order probablity $i.e.,$ the $q
 \rightarrow 1 $ value:
\be
P_n^{(0)}=\lim \frac{1}{\cal Z}  e^{ - \ds \frac{x}{2}(\, [n]-[n+1]\, )}
= \frac{1}{\cal Z}_0 e^{ - \ds \frac{x}{2} (-1)^{n+1}  } .
\label{24}
\ee
where ${\cal Z}_0$  is given by Eq.(\ref{2}). This contains and reproduces
the classical values for $P^{(0)}_0, P^{(0)}_1$ but yet $n$ is not
 restricted to
$0,1$ and so repeats the same pair of values for various values of $n$. In 
other words Pauli principle has to be imposed explicitly by hand. The 
distribution function is
\be
f_q =\sum_{n=0}^{\infty}  \frac{1}{2} \left ( [n]-[n+1] \right ) P_n\; , 
\label{25}
\ee
which cannot be evaluated in a closed form, just as in the case of
$q$-deformed boson oscillators \cite{pns} because of the occurrence of $[n]$ 
here. Of the two methods known in the literature, we shall choose and
 employ the method of iteration suggested by Song et al \cite{Songetal} and
 developed fully in ref. \cite{pns}. As  a first iteration, one may
  approximate the  
probability function $P_n$ in Eq.(\ref{25}) by the lowest order value, 
$P^{(0)}_n$, given by Eq.(\ref{24}) thus: 
\be
f_q =\sum_{n=0}^{\infty} \frac{1}{2} \left (  [n]-[n+1] \right )P^{(0)}_n.
\label{26}
\ee
We accordingly obtain the first order iteration distribution function
 for the 
generalized $q$-deformed Fermions:
\be
f_q =\sum_0^{\infty} \frac{1}{4 \cosh \frac{x}{2}}\, e^{- \ds \frac{x}{2}  }
(-1)^{n+1}   \; \frac{q}{1+q^2}\left \{ q^{-n}(1- q^{-1}) - (-q)^n
 (1+q) \right 
\}.
\label{27}
\ee
It turns out that this does not lead to the correct high temperature limit
 for 
the internal energy, nor does it lead to the correct $q \rightarrow 1$ limit. 
This can be demonstrated as follows. If we isolate the even and odd numbers
 in 
the sum in Eq.(\ref{27}), we obtain
\br
f_q &=&\frac{1}{4 \cosh \frac{x}{2}}\, \frac{q}{1+q^2} \left ( 
\sum_{n= even}^{\infty} e^{\ds \frac{x}{2}} \left \{
(1-q^{-1})q^{-n} - (1+q) q^n \right \} \right.   \nonumber \\
&+&  \left.   \sum_{n= odd} ^{\infty}e^{-\ds \frac{x}{2}} \left \{
(1-q^{-1})q^{-n}+ (1+q) q^n \right \}  \right )
\er 
Evaluating the sums we obtain
\be
f_q = \frac{1}{2} \; \frac{q}{q^2-1} \tanh \frac{x}{2}.
\ee
The internal energy is
\be
U=N \hbar \o f_q = \frac{1}{2}  N \hbar \o \frac{q}{q^2-1} \tanh x/2.
\ee
If we examine the high temperature limit, $ x <<1$ we find that the
internal energy indeed vanishes as $T \rightarrow \infty$ for any
 value of $q$. The low temperature limit however, as $x >> 1$,  becomes
 \be
U_{T \rightarrow 0}= \frac{1}{2} N \hbar \o \frac{q}{q^2-1},
\ee
which depends on $q$ and does not agree with the classical limit.
 It thus appears that the low temperature behavior does not agree with
  the undeformed 
Fermion theory,  contrary to what we would expect \cite{Birman}. Furthermore
it is singular at $q=1$. In other words the internal energy as well as the 
distribution function do not possess the correct classical limits.
The cause of this problem can be traced to the fact that even in the first
order iteration, the sum extends over
$n=0, 1, \cdots, \infty $;  on the other 
hand, we expect only $n=0,1$ to contribute to the sum because in the lowest
order, the probability function $P^{(0)}_n$ is expected to be non-zero
 only for $n=0,1$. To 
resolve this problem we recall that Pauli exclusion principle is not
 contained in the definition of the probability $P^{(0)}_n$ in Eq.(\ref{24})
  but has to
be imposed explicitly. Instead we ought to use the weak exclusion
 principle, in the operator sense. Recall also that according to
  Eq.(\ref{18}) the operator $(b^{\dagger})^n$, in the limit
   $q \rightarrow 1$, is non-zero for $n=0,1$ and
vanishes for $n>1$ only in the weak operator sense. Analogously we must 
define the weak operator relation for the zeroth order probability as
\be
P^{(0)}_n\;  = \, P^{(0)}_0\, |0> + P^{(0)}_1 \, |1>
\ee
We can also interpret this as
\be
P^{(0)}_n = \d_{n0} P^{(0)}_0+ \d_{n1}P^{(0)}_1,
\ee
in which case we obtain
\be          
P^{(0)}_n \, |n> = \frac{1}{2 \cosh x/2} \left \{ \d_{n0}e^{\ds \frac{x}{2}}
+ \d_{n1}e^{-\ds \frac{x}{2}}   \right \} \, \, |n>.
\ee
Only with this interpretation, with exclusion principle in the weak sense,
 the $q$-deformed generalized Fermion theory will reduce to the standard
  Fermion theory in the $q\rightarrow 1$ limit in a self-consistent manner. 

The correct first order iteration of the distribution is thus given
by Eq.(\ref{27}) but with the sum over only $n=0,1$ and hence it reduces
to the expression given by  Eq.(\ref{7}). We have already shown that
this has the corrrect low temperature limit and the correct
high temperature limit.

In summary, we have developed the theory of the statistical mechanics of
$q$-deformed generalized Fermions, based on the algebra of creation and
annihilation operators proposed in
 Ref.\cite{Partha}. We derived the approximate form for the distribution
 function based on the first order iteration. We then found
 that this did not possess the correct low temperature limit, nor did it
 reproduce the classical result when
 $q \rightarrow 1$. By introducing the weak exclusion principle into the
 theory for the probability function, we are able to show that this leads to
 all the desired properties. This also has the desirable feature that at
 very high temperature as well as at very low temperature, the theory
 reduces to that of undeformed Fermions. As a corollary we find that although
 the algebra of the creation and annihilation operators for arbitrary
 deformation is quite different from, and more general than, that of the
 standard undeformed Fermions, the statistical mechanics of 
 $q$-deformed generalized Fermions with weak exclusion principle reduces
 to the statistical mechanics of undeformed Fermions in the first
 order of iteration.

\vspace{.2in}

\noindent {\bf Acknowledgments}

\vspace{.2in}

 I am thankful to A. Lavagno for valuable discussions, both while I
 was visiting Politecnico di Torino, Torino, Italy in 1997 and subsequently by
 correspondence, and for drawing my attention to some work on the subject
 of $q$-oscillators. I thank R. Ramachandran and G. Rajasekaran,
for hospitality at the Institute of Mathematical Sciences, Chennai and
R. Parthasarathy for many fruitful discussions.

\end{document}